\documentclass[aps,amsfonts,amssymb,twocolumn,amsmath,superscriptaddress,prl]{revtex4-2}[12pt]

\usepackage{graphicx}% Include figure files
\usepackage{dcolumn}% Align table columns on decimal point
\usepackage{bm}% bold math
\usepackage{xcolor}
\usepackage[pdftex,colorlinks=true,linkcolor=blue,citecolor=blue,urlcolor=blue,filecolor=blue,breaklinks=true]{hyperref} %clickable links in refs
\usepackage{url}
\usepackage{braket}

\usepackage{setspace}

\usepackage{textcomp}
\usepackage{gensymb}
\usepackage{mathtools}
\usepackage{cleveref}
\usepackage{siunitx}
\usepackage[draft]{changes}

\DeclareSIUnit{\angstrom}{\text {Å}}
\DeclareSIUnit{\gauss}{\text {G}}
\usepackage{float}
\usepackage{comment}
\usepackage{verbatim}
\usepackage{gensymb}
\usepackage{array}
\usepackage[version=4]{mhchem}
\usepackage{manfnt}

\definechangesauthor[name={Rishi Bhandia},color=orange]{RB}

\begin{document}

\title{Energy and momentum relaxation through the Curie temperature in an itinerant ferromagnet}
\author{Rishi Bhandia}
\thanks{rb3926@nyu.edu}
\affiliation{William H. Miller III Department of Department of Physics and Astronomy, The Johns Hopkins University, Baltimore, Maryland 21218, USA}

\author{Tim Priessnitz}
\affiliation{Max Planck Institute for Solid State Research, Heisenbergstraße 1, 70569 Stuttgart, Germany}

\author{Jiahao Liang}
\affiliation{William H. Miller III Department of Department of Physics and Astronomy, The Johns Hopkins University, Baltimore, Maryland 21218, USA}

\author{Ksenia S. Rabinovich}
\affiliation{Max Planck Institute for Solid State Research, Heisenbergstraße 1, 70569 Stuttgart, Germany}

\author{Ralph Romero III}
\affiliation{William H. Miller III Department of Department of Physics and Astronomy, The Johns Hopkins University, Baltimore, Maryland 21218, USA}

\author{Kota Katsumi}
\affiliation{William H. Miller III Department of Department of Physics and Astronomy, The Johns Hopkins University, Baltimore, Maryland 21218, USA}

\author{Thi Thu Huong Tran}
\affiliation{Max Planck Institute for Solid State Research, Heisenbergstraße 1, 70569 Stuttgart, Germany}

\author{Georg Christiani}
\affiliation{Max Planck Institute for Solid State Research, Heisenbergstraße 1, 70569 Stuttgart, Germany}

\author{Gennady Logvenov}
\affiliation{Max Planck Institute for Solid State Research, Heisenbergstraße 1, 70569 Stuttgart, Germany}

\author{Bernhard Keimer}
\affiliation{Max Planck Institute for Solid State Research, Heisenbergstraße 1, 70569 Stuttgart, Germany}

\author{N. P. Armitage}
\thanks{npa@jhu.edu}
\affiliation{William H. Miller III Department of Department of Physics and Astronomy, The Johns Hopkins University, Baltimore, Maryland 21218, USA}

\begin{abstract}
In this work, we combine conventional linear response time-domain THz spectroscopy with non-linear THz-pump THz-probe techniques to study metallic strained thin films of $\mathrm{Ca}_2\mathrm{RuO}_4$, which undergo a transition into a ferromagnetic state at \qty{10}{\K}.  Such measurements allowing us to independently measure momentum and energy relaxation rates. We find that while the momentum relaxation rate decreases significantly at the ferromagnetic transition, the energy relaxation rate remains unaffected by the emergence of magnetic order. This shows that the dominant changes to scattering across the transition correspond to scatterings that relax momentum without relaxing energy.  It is consistent with a scenario where energy is not carried off by coupling to collective magnetic degrees of freedom. Instead, the principal channel for energy relaxation remains the conventional one e.g. coupling to acoustic phonons. This observation validates the approximation used in the conventional understanding of resistive anomalies of ferromagnets across the Curie temperature, which due to critical slowing down, spin fluctuations can be treated as effectively static and scattering off of them elastic. This scenario can likely be extended to resistive anomalies at other phase transitions to charge- and spin-density wave states in kagome metals or pnictide systems.

\end{abstract}

\date{\today}
\maketitle

% Anomalous Bloch-Grüneisen

%\section{Introduction}

Magnetism and metallicity are some of the oldest aspects of materials studied by humans. However, our understanding of their interplay remains incomplete~\cite{gerlach1932modification,craig1967transport,moriyaRecentProgressTheory1979,kleinAnomalousSpinScattering1996,santiagoItinerantMagneticMetals2017}.  It is common to observe an anomaly in the resistivity of metallic ferromagnets, $\rho(T)$, as they are tuned through a ferromagnetic phase transition. In the ferromagnetic phase, the resistivity of the metal then decreases with a steeper slope, corresponding to the suppression of momentum-relaxing scattering channels~\cite{gerlach1932modification,craig1967transport,wangSubterahertzMomentumDrag2020,kleinAnomalousSpinScattering1996}.  The origin of the suppressed scattering is unclear, but is generally believed to arise from the quenching of scattering of electrons off of spin fluctuations that are correlated over some length scale $\xi$.   Long ago, de Gennes and Friedel pointed out that due to critical slowing down, spin fluctuations may be approximated as effectively static near the Curie temperature~\cite{de1958anomalies}. Building upon this work, Fisher and Langer showed~\cite{fisher1968resistive} that short-range spin fluctuations are the principal contributors to the resistivity and internal energy of the system, suggesting that the temperature derivative of the resistivity and heat capacity should follow the same temperature dependence.  Some experiments have shown rough agreement with theory~\cite{zumsteg1970electrical,shacklette1974specific}, while others have noted disagreements~\cite{kleinAnomalousSpinScattering1996}.  It has been argued a inclusion of inelastic scattering processes is important to provide a full theoretical description of experimental results~\cite{geldart1977effect}.

Transition metal oxides with $4d$ and $5d$ valence electrons can have strong onsite Coulomb repulsion, crystal field splittings, and spin-orbit coupling that can reach comparable energy scales of \qtyrange{0.2}{0.4}{\eV} ~\cite{takayamaSpinOrbitEntangledElectronic2021}.  This gives opportunity for tuning of the various competing interactions and the emergence of different magnetic states. %Moreover, $d$-electrons have properties intermediate between the extended long-range wave-functions of $p$ and $s$-orbitals and the localized orbitals of $f$-electrons. This leads to narrow electronic bandwidths of \qtyrange{1}{2}{\eV}, small compared to bandwidths of \qtyrange{5}{15}{\eV} typical for metals~\cite{raoTransitionMetalOxides1989}.
Perovskite ruthenates, with their $4d$-orbitals, are a paradigmatic example with this tunability. The extended nature of $4d$-orbitals in ruthenates leads to enhanced hybridization with oxygen $2p$ orbitals, facilitating a delicate balance between itinerant and localized electrons ~\cite{caoAntiferromagneticInsulatorFerromagnetic1999}. This makes ruthenates highly susceptible to external perturbations such as pressure, chemical doping, and epitaxial strain, allowing for a broad spectrum of phases. For instance, compounds like $\mathrm{Sr}_2\mathrm{RuO}_4$ display unconventional superconductivity~\cite{maenoSuperconductivityLayeredPerovskite1994, maenoStillMysteryAll2024, armitageSuperconductivityMysteryTurns2019} that is highly sensitive to pressure and strain~\cite{grinenkoSplitSuperconductingTimereversal2021,steppkeStrongPeakTc2017,jerzembeckSuperconductivitySr2RuO4Caxis2022}.

$\mathrm{Ca}_2\mathrm{RuO}_4$ is another example.  Crystals exhibit Mott insulating behavior and antiferromagnetism~\cite{nakatsujiCa2RuONew1997}, and host a dramatic structural transition when heated past roughly \qty{360}{\kelvin} that changes its electronic properties to that of a metal~\cite{alexanderDestructionMottInsulating1999,caoGroundstateInstabilityMott2000}. This transition corresponds to a change in lattice volume of \qty{1.3}{\percent}. The transition is so severe that single crystals shatter, making growth of large single crystals difficult. The application of pressure along the c-axis changes the ground state of $\mathrm{Ca}_2 \mathrm{RuO}_4$ from an antiferromagnetic insulator to a ferromagnetic metal~\cite{fangMagneticPhaseDiagram2001,friedtStructuralMagneticAspects2001,nakamuraMottInsulatorFerromagnetic2002, taniguchiAnisotropicUniaxialPressure2013}. This electronic configuration favors ferromagnetic correlations at low temperatures, with spins aligned along the a-axis~\cite{nakamuraMottInsulatorFerromagnetic2002,steffensHighpressureDiffractionStudies2005}. The magnetic moment per Ru is a reduced relative to its full moment, $0.4\ \mu_B$~\cite{nakamuraMottInsulatorFerromagnetic2002}, indicating itinerant ferromagnetism. This structural transition induces changes in the orbital, electronic, and spin degrees of freedom and further illustrates their intricate interplay and their sensitivity to lattice distortions.

The complexity and tunability of $\mathrm{Ca}_2\mathrm{RuO}_4$ has led to great interest in exploring its engineered ``designer'' phases via epitaxial growth on mismatched substrates and the stabilization of meta-stable phases~\cite{schlomThinFilmApproach2008, rameshCreatingEmergentPhenomena2019,ahnDesigningControllingProperties2021,miaoItinerantFerromagnetismGeometrically2012,dietlSynthesisElectronicOrdering2018}.  Additionally, it has been found that moderate applied current can drive the metal-insulator transition. This occurs with applied voltages much lower than the insulating gap~\cite{cirilloEmergenceMetallicMetastable2019, fursichRamanScatteringCurrentstabilized2019, jenniEvidenceCurrentinducedPhase2020, nakamuraElectricfieldinducedMetalMaintained2013, okazakiCurrentInducedGapSuppression2013}. It has also been found that this transformation can be induced via excitation with ultra-short \qty{800}{\nm} (\qty{1.5}{\eV}) pulses~\cite{vermaPicosecondVolumeExpansion2024}. These observations underline the importance of understanding the interplay between $\mathrm{Ca}_2\mathrm{RuO}_4$'s various degrees of freedom in non-equilibrium conditions.

\begin{figure}[t]
\begin{center}
	\includegraphics[width = 0.9\linewidth]{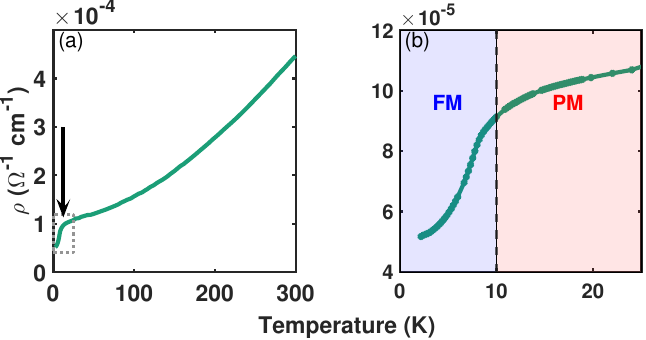}
\end{center}
\caption{(a) The resistivity of the \qty{35.6}{\nm} thin film of $\mathrm{Ca}_2 \mathrm{RuO}_4$ measured in a 4-probe geometry with an excitation current of \qty{10}{\uA}. (b) The region enclosed by the grey box in (a) enlarged to emphasize impact of ferromagnetic  transition on the resistivity}\label{fig:CRODCresistivity}
\end{figure}

The $\mathrm{Ca}_2\mathrm{RuO}_4$ films studied in this work are under compressive strain and are observed to exhibit a ferromagnetic transition at \qty{10}{\K}. Like many itinerant ferromagnets they show a resistive anomaly below the Curie temperature (Fig.~\ref{fig:CRODCresistivity}) where the resistivity shows a steeper temperature dependence below the ferromagnetic transition compared to above it.  In this study we combine time-domain terahertz (TDTS) with non-linear THz-pump THz-probe measurements~\cite{mahmoodObservationMarginalFermi2021,katsumiRevealingNovelAspects2024,barbalasEnergyRelaxationDynamics2023} of strained thin films of $\mathrm{Ca}_2\mathrm{RuO}_4$ to separately measure its momentum and energy relaxation rates.  Conventional linear response TDTS measurements have shed light on the details of momentum transport in other correlated ruthenates~\cite{wangSubterahertzMomentumDrag2020,wangSeparatedTransportRelaxation2021}.  We find that momentum relaxation is reduced dramatically through the ferromagnetic transition while energy relaxation is insensitive to the onset of magnetic order.  Although one may have expected that coupling to emergent collective magnetic degrees of freedom could remove energy from the electronic system this appears not to be the case.  This observation supports the approximation commonly proposed for understanding resistive anomalies in ferromagnets, where, due to critical slowing down, spin fluctuations are treated as effectively static as the system approaches the Curie temperature, making the scattering off these fluctuations elastic.   Our data indicates that such elastic scattering is the dominant scattering that relaxes current.  Energy relaxation presumably proceeds through the acoustic phonons which are the conventionally dominant channel for energy relaxation in metals.

%\section{Experimental Details}

We studied a \qty{35.6}{\nm} thick thin film of $\mathrm{Ca}_2\mathrm{RuO}_4$ grown on a $\mathrm{LaSrAlO}_4$ (LSAO) (001) substrate.  Compressive strain of the substrate stabilized the metallic phase of $\mathrm{Ca}_2\mathrm{RuO}_4$~\cite{dietlTailoringElectronicProperties2018}. The film was grown using an off-axis sputtering setup described in Ref~\cite{rabinovichEpitaxialGrowthStoichiometry2024}. The deposition process utilized  $\mathrm{Ca}_2\mathrm{RuO}_4$ and ruthenium targets, applying a sputtering power ratio of 3:1 W. The film was deposited in a mixed argon-oxygen (1:1) atmosphere at a total pressure of \qty{0.1}{\milli \bar} and at a substrate temperature of \qty{650}{\degreeCelsius}. This strained film shows a ferromagnetic (FM) transition at \qty{10}{\K} with an easy axis in plane.  We conducted non-linear THz-pump THz-probe measurements in a nonlinear THz 2DCS spectrometer, in the now-standard collinear geometry~\cite{mahmoodObservationMarginalFermi2021,katsumiRevealingNovelAspects2024,bhandiaAnomalousElectronicEnergy2024a,barbalasEnergyRelaxationDynamics2023}. The laser pulse was split into three parts, using two parts of it to pump two $\mathrm{LiNbO}_3$ THz sources, and the other for electro-optic sampling. Displayed data was taken with a pump pulse with a peak amplitude of \qty{80}{\kV \per \cm} and a probe pulse with a peak amplitude of \qty{10}{\kV \per \cm}.  Experiments were also done with weaker pump pulses and although had larger noise gave data largely consistent with what is shown here.  These experiments were conducted in a transmission geometry with a crossed polarization of pump and probe. The fluence for each pulse was controlled by a pair of wire-grid polarizers. At the detection crystal, a wire grid polarizer was used to minimize the detection of the pump pulse. To measure the non-linear response, a differential chopping scheme was used to measure the transmitted pulses, $E_{A}$ and $E_B$ separately, and then pulses $E_{AB}$ transmitted together. From this, the non-linear component of the response can be calculated via subtraction, $E_{NL} = E_{AB} - E_A - E_B$.  This spectrometer was integrated into a cold-finger helium flow cryostat, with the sample heat sunk to the cold finger using Apeizon N grease and copper tape. The pump and probe field strengths were characterized using a \qty{0.5}{\mm} GaP crystal placed at the sample location before switching to the cryostat.

\begin{figure}[t]
	\begin{center}
		\includegraphics[width = 0.95\linewidth]{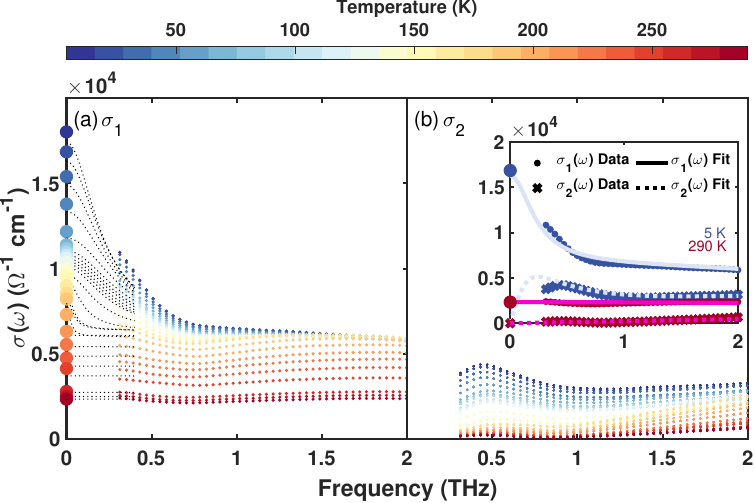}
	\end{center}
		 \caption{The THz optical conductivity of $\mathrm{Ca}_2 \mathrm{RuO}_4$.  (a) and (b) depict the real and imaginary parts respectively, with the dotted black lines in (a) showing the Drude extrapolation connecting the DC and THz conductivity. The inset shows examples of the Drude fits to THz conductivity at 5 K and 290 K respectively.}\label{fig:CROlinearResponse}
\end{figure}

%\section{Analysis and Results}
The DC resistivity of the film measured in a 4-contact geometry is presented in Fig.~\ref{fig:CRODCresistivity}. Over much of the temperature range the resistivity shows the expected metallic temperature dependence, decreasing with decreasing temperature. When the sample is cooled through the Curie temperature, \qty{10}{\K}, the resistivity decreased by an additional factor of 2 over a small temperature range.  This feature and the quenched magnetic moment of the Ru ions~\cite{dietlSynthesisElectronicOrdering2018}, are consistent with an itinerant ferromagnetic ground state, similar to the pressure-induced metallic phase of $\mathrm{Ca}_2 \mathrm{RuO}_4$~\cite{nakamuraMottInsulatorFerromagnetic2002}.

\begin{figure}
	\begin{center}
		\includegraphics[width = 0.9\linewidth]{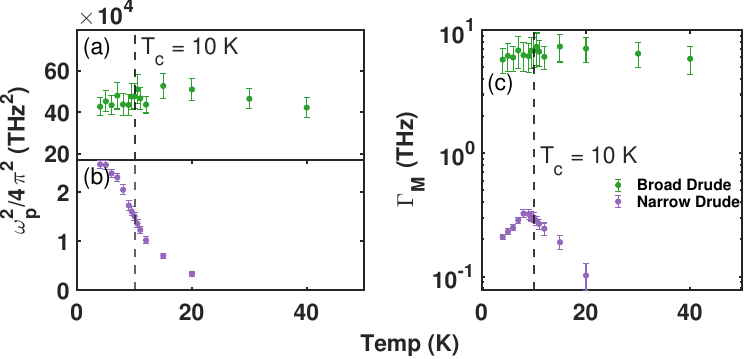}
	\end{center}
		 \caption{The results of the Drude fits to the THz conductivity. (a) The spectral weight of the broad Drude term. (b)  The spectral weight of narrow Drude term. (c)  The momentum relaxation rate corresponding to the narrow and broad Drude terms respectively}\label{fig:croDrudeFits}
\end{figure}

The real and imaginary parts of the optical conductivity are presented in Fig.~\ref{fig:CROlinearResponse}(a) and (b). The DC conductivity is plotted as a zero frequency point in Fig.~\ref{fig:CROlinearResponse}(a), with the dotted lines representing extrapolations of the optical conductivity to zero frequency, showing the agreement between the DC conductivity and the measured optical conductivity.  Examining the temperature dependence of the optical conductivity, it is apparent that, consistent with prior measurements on other ruthenates~\cite{wangSeparatedTransportRelaxation2021, wangSubterahertzMomentumDrag2020}, the complex conductivity can be well modeled as a sum of two Drude terms at low temperatures,

\begin{align}
	\sigma(\omega) = \epsilon_0 \left[\frac{-\omega_{1,p}^2}{i\omega - 2 \pi \Gamma_{1,M}} + \frac{-\omega_{2,p}^2}{i\omega - 2 \pi \Gamma_{2,M}} - i(\epsilon_\infty - 1)\omega\right],
\end{align} 
crossing over into a regime characterized by a single, broad Drude at higher temperatures. Here, $\omega_{p,1}$ and $\omega_{p,2}$ are the Drude plasma frequencies, $\Gamma_{1,M}$ and $\Gamma_{2,M}$ are the current relaxation rates, and $\epsilon_\infty$ is the high-frequency dielectric constant accounting for the effects of higher band inter-band transitions on the low-frequency dielectric constant. It is clear that THz-range optical conductivity is consistent with the measured DC resistivity. This is reflected in Fig.~\ref{fig:croDrudeFits}(a) and (b), where the spectral weight corresponding to the narrow Drude, increases dramatically when cooling through the Curie temperature while the broad Drude term is relatively unchanged. This indicates that formation of the narrow Drude component is the origin of the suppression of DC resistivity observed at the Curie temperature. Example fits for the optical conductivity at \qty{5}{\K} and \qty{290}{\K} are shown in the inset of Fig.~\ref{fig:CROlinearResponse}(b). The scattering rates corresponding to the two Drude components, $\Gamma_{1,M}$ and $\Gamma_{2,M}$, are plotted as a function of temperature in Fig.~\ref{fig:CROlinearResponse}(c). Strictly, these rates are actually connected to different channels for current relaxation~\cite{lavasaniWiedemannFranzLawFermi2019}, but for materials with sufficiently isotropic dispersion, they are equivalent to the momentum relaxation rate. It is interesting to note in Figs.~\ref{fig:croDrudeFits}(a) and (c) that the broad Drude term does not show any dependence on temperature with a momentum relaxation rate at least two orders of magnitude larger than the narrow Drude term. Although we assume that is that the broad feature comes from a Drude transport channel, we cannot exclude the fact that it arises from very low energy inter-band transitions (\qtyrange{1}{2}{\meV}) that may arise from rotations the octahedra~\cite{dang2015band}. In either case our below interpretation is unchanged.

\begin{figure}
	\begin{center}
		\includegraphics[width = 0.9\linewidth]{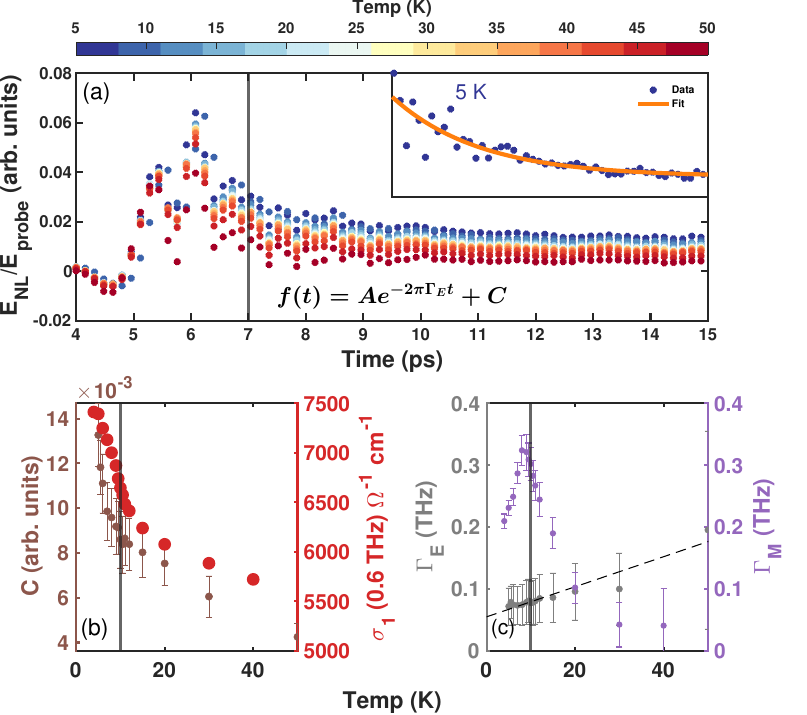}
	\end{center}
		 \caption{Non-linear THz-pump THz-probe data. (a) Temperature-dependent THz-pump THz-probe time traces. The inset shows an example exponential fit. To avoid the artifact arising from overlap of pump and probe pulses, only the data from \qty{7}{\ps} (horizontal grey line) onwards is fit with the exponential model.  (b) Comparison of the $\sigma_1\left(\omega = \qty{0.6}{\THz}\right)$ and the DC component of the exponential fit. (c) Comparison of the energy relaxation rate $\Gamma_{E}$ and the momentum relaxation rate from the narrow Drude $\Gamma_{M,2}.$ }\label{fig:PumpProbeTimeTrace}
\end{figure}

In Fig.~\ref{fig:PumpProbeTimeTrace}(a) we show pump-probe time traces measured at different temperatures.  In keeping with our previous work and theory~\cite{confortiDerivationThirdorderNonlinear2012}, our expectation is that the NL THz response in a metal comes purely from pump-probe processes.  The observation exponential decay at times longer than our THz pulses is consistent with this expectation.  At short times (\qtyrange{5}{6.5}{\ps}), the oscillating signal derives from pump-probe processes where the probe and pump pulses swap roles when probe precedes the pump in time~\cite{barbalasEnergyRelaxationDynamics2023}.  The data shows fast initial decay and then a long time tail, which we believe is indicative of two time scales, the longer of which is longer than the experimental time window. The dynamics speeds up at higher temperature, and by \qty{50}{\K} the measured decay is nearly impossible to distinguish above the noise floor.  To interpret the data, we fit the time traces with an exponential decay plus a constant to capture the long-lived timescale,
\begin{align}
E_\text{NL}(t) \propto A e^{-2\pi \Gamma_E t} + C.
\label{eq:pumpProbeFit}
\end{align}
Here, $A$ represents the amplitude of the rise driven by the pump pulse, $\Gamma_E$ is the decay rate and $C$ is the amplitude of the long-lived signal.  We fit only to times longer than \qty{7}{\ps}, where the clearly decaying signal is observed. An example fit, along with its two individual components, is shown in the inset of Fig.~\ref{fig:PumpProbeTimeTrace}(a). Consistent with prior measurements on metallic ruthenates~\cite{barbalasEnergyRelaxationDynamics2023}, we believe that the shorter of the two timescales correspond to the decay of energy from the electronic system. The longer timescale then may correspond to be the much slower diffusion of energy out of the excited region~\cite{kawasakiThermalDiffusivity2021} or from slow relaxation of directly excited very low energy phonons ($\approx$ \qty{2}{\meV}) that may arise from rotations of the octahedra in the ideal perovskite structure~\cite{Hameed2024}.

In Fig.~\ref{fig:PumpProbeTimeTrace}(b)-(c), we show the relevant fit parameters as a function of temperature. The long-lived component of the pump-probe time trace corresponding the constant component in Eq.~\ref{eq:pumpProbeFit} is plotted in Fig~\ref{fig:PumpProbeTimeTrace}(b). As can be deduced from the time traces shown, this long-lived component increases dramatically through the magnetic transition, consistent with more heat being dissipated at low temperatures as the sample becomes more conductive.   Comparing this to the dissipative part of the conductivity $\sigma_1 \left(\omega = \qty{0.6}{\THz}\right)$ in Fig.~\ref{fig:PumpProbeTimeTrace}(a), we can see the increase in this long-lived state is correlated to the rise in conductivity, suggesting that this long-lived state is connected to heating.  In Fig.~\ref{fig:PumpProbeTimeTrace}(c) the energy relaxation rate decreases as the sample is cooled, consistent with the expected low temperature behavior~\cite{allenTheoryThermalRelaxation1987,gloriosoJouleHeatingBad2022} for energy relaxation in a metal at temperatures below the Bloch-Gr{\"u}neisen temperature.  Quite notably the energy relaxation rate does not change through the ferromagnetic transition.  This is in stark contrast to the momentum relaxation rate, which changes strongly through the Curie temperature and is the source of the resistive drop in Fig.~\ref{fig:CRODCresistivity}.

% Similarity to CDWs.  General mechanism

% Phonons directly excited.

%\begin{figure}
%	\begin{center}
%		\includegraphics[width = 0.9\linewidth]{Figure5.pdf}
%	\end{center}
%		 \caption{Comparison of the fit results from the linear and non-linear THz data. (a) Comparison of the $\sigma_1\left(\omega = \qty{0.6}{\THz}\right)$ and the DC component of the exponential fit. (b) Comparison of the energy relaxation rate $\Gamma_{E}$ and the momentum relaxation rate from the narrow Drude $\Gamma_{M,2}.$ }\label{fig:FitParameterComparision}
%\end{figure}

%\section{Conclusion}
In this work we have used linear and non-linear THz spectroscopy to probe the electronic energy and momentum relaxation of $\mathrm{Ca}_2 \mathrm{RuO}_4$. The linear optical response spectra are well described by two Drude components, a narrow term that arises at low temperatures, and a broad term that persists to the highest temperatures measured. This behavior is qualitatively similar to ruthenates that have been studied before, but here this narrow Drude appears at the ferromagnetic transition. The energy relaxation rate shows a temperature dependence that is monotonically increasing as temperature increases, qualitatively consistent with the $T^3$ behavior expected at temperatures below the Bloch-Gr{\"u}neisen temperature. In contrast to the momentum relaxation rate, this rate is insensitive to the onset of magnetic order.  This implies that the scattering processes that change across the transition primarily relax momentum without relaxing energy. It gives evidence that energy is not dissipated via coupling to collective magnetic degrees of freedom, and instead, the main energy relaxation mechanism is the conventional one e.g. coupling to acoustic phonons. This observation validates the approximation used in the standard understanding of resistive anomalies in ferromagnets, where, due to critical slowing down, spin fluctuations can be considered effectively static near the Curie temperature, with scattering off these fluctuations being almost entirely elastic. This scenario can likely be extended to resistive anomalies at classic charge-density wave transitions like in NbSe$_3$~\cite{naito1982electrical}, newer kagome metals with charge-density waves~\cite{cao2023competing}, and spin-density wave systems in e.g. pnictides~\cite{tanatar2010pseudogap}.

We thank Chris Lygouras and Vincent Morano for helpful discussions on the DC resistivity measurements.
Work at JHU was supported by the ARO MURI ``Implementation of axion electrodynamics in topological films and device'' W911NF2020166.  Instrumentation development at JHU was supported by the Gordon and Betty Moore Foundation EPiQS Initiative Grant GBMF-9454. Sample synthesis and characterization at MPI FKF was supported by funding from the European Research Council (ERC) under Advanced Grant no. 101141844 (SpecTera).

\bibliography{CRO_firstdraft}

\end{document}